\documentclass{kluwer}

\newdisplay{guess}{Conjecture}
\usepackage{graphicx}

\newcommand{\vu}{\vec{u}}
\newcommand{\vpsi}{\vec{\psi}}
\def\Ln{\mathop{\hbox{ln}}\nolimits}
\def\Div{\mathop{\hbox{div}}\nolimits}
\newcommand{\lp}{ \left(}
\newcommand{\rp}{ \right)}
\newcommand{\na}{ \vec{\nabla} }

\begin{document}                                                                                   
%\begin{article}
\begin{opening}         
\title{Stochastic Excitation of Gravity Waves \\
by Overshooting Convection in Solar-Type Stars}
\author{Boris \surname{Dintrans}$^1$}  
\author{Axel \surname{Brandenburg}$^2$}  
\author{\AA ke \surname{Nordlund}$^3$}  
\author{Robert F. \surname{Stein}$^4$}
\runningauthor{Boris Dintrans, Axel Brandenburg, \AA ke Nordlund and
Robert F. Stein}
\runningtitle{Stochastic excitation of gravity waves by overshooting
convection}
\institute{$^1$ Observatoire Midi-Pyr\'en\'ees, CNRS UMR5572, 14 avenue 
Edouard Belin, F-31400 Toulouse, France}
\institute{$^2$ NORDITA, Blegdamsvej 17, DK-2100 Copenhagen \O, Denmark}
\institute{$^3$ Astronomical Observatory, Juliane Maries Vej 30, DK-2100 
Copenhagen \O, Denmark}
\institute{$^4$ Department of Physics and Astronomy, Michigan State
University, East Lansing, MI48824, USA}
%\date{\today}

\begin{abstract}
The excitation of gravity waves by penetrative convective plumes is
investigated using 2D direct simulations of compressible convection.
The oscillation field is measured by a new technique based on the
projection of our simulation data onto the theoretical $g$-modes solutions
of the associated linear eigenvalue problem. This allows us to determine
both the excited modes and their corresponding amplitudes accurately.
\end{abstract}
\end{opening}           

\section{Introduction}  

Two-dimensional simulations of compressible convection have shown that
it is possible to excite internal gravity waves (IGW) in radiative zones
of solar-type stars from the downward penetrating plumes (Hurlburt et al.,
 \citeyear{Hur86}, hereafter HTM86; Hurlburt et
al., \citeyear{Hur94}; Kiraga and Jahn, \citeyear{Kir95}).  However,
detecting IGW with confidence is challenging given the stochastic nature
of the excitation mechanism.

We propose here a new detection method which allows us to measure
rigorously both the spectrum and amplitude of excited $g$-modes. This
method is first applied to the $g$-mode oscillations of an isothermal
atmosphere and then, to IGW generated in a 2D-simulation of a convective
zone embedded between two stable ones.

\section{Detecting $g$-modes using the anelastic subspace}

In hydrosimulations, wave fields are commonly measured using two main
methods: {\it (i)} the simplest one consists in recording the vertical
velocity at a fixed point and then performing the Fourier transform of
the sequence (see, e.g., HTM86); {\it (ii)} a more complicated method
consists in taking two Fourier transforms, in space and time, of the
vertical mass flux (Stein and Nordlund, \citeyear{Stein90}). However, these
two methods are not well adapted to detect IGW in our problem because
the Fourier transforms are calculated over the {\it whole} simulation
while IGW are {\it stochastically} excited by penetrating plumes.

Our new detection method takes into account the random nature of
this excitation. Indeed, it is based first, on projections of the
simulated velocity field $\vec{v}(k,z,t)$ onto the anelastic eigenvectors
$\vpsi_{kn}(z)$ as

\begin{equation}
\vec{v} (k,z,t) = \sum_{n=0}^\infty <\vpsi_{kn},\vec{v}> \vpsi_{kn} (z) =
\sum_{n=0}^\infty A_{kn} (t) \vpsi_{kn}(z), 
\label{projec}
\end{equation}
and second, on time-frequency diagrams of the complex coefficients
$A_{kn}(t)$. As a consequence, the immediate spectrum (the set
of frequencies $\omega$) and amplitudes (defined as $|A_{kn}|$) of
stochastically excited $g$-modes are reached and not only their ``mean''
values over the whole simulation. It is instructive to consider the
simplest possible case, that is, the propagation of a single gravity mode
with horizontal wavenumber $k$ through the computation domain: applying
eq. (\ref{projec}) leads in this case to a projection coefficient $A_{kn}
(t) \propto \exp(i\omega_{kn} t)$, where $\omega_{kn}$ denotes the
frequency of the anelastic eigenmode of degree $k$ and order $n$.

\section{Results}

\begin{figure}
\centerline{\includegraphics[width=0.5\textwidth]{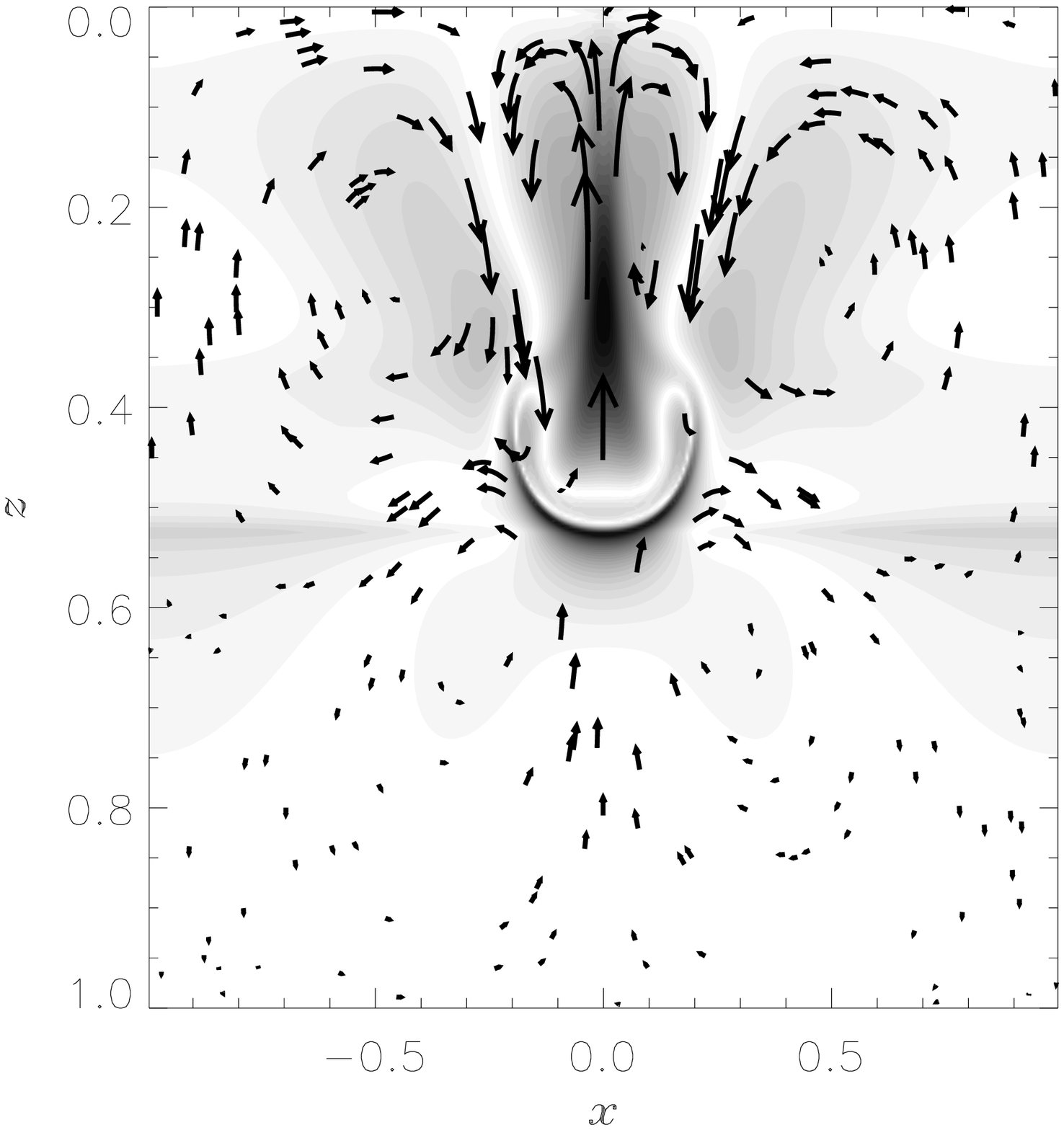}
\includegraphics[width=0.5\textwidth]{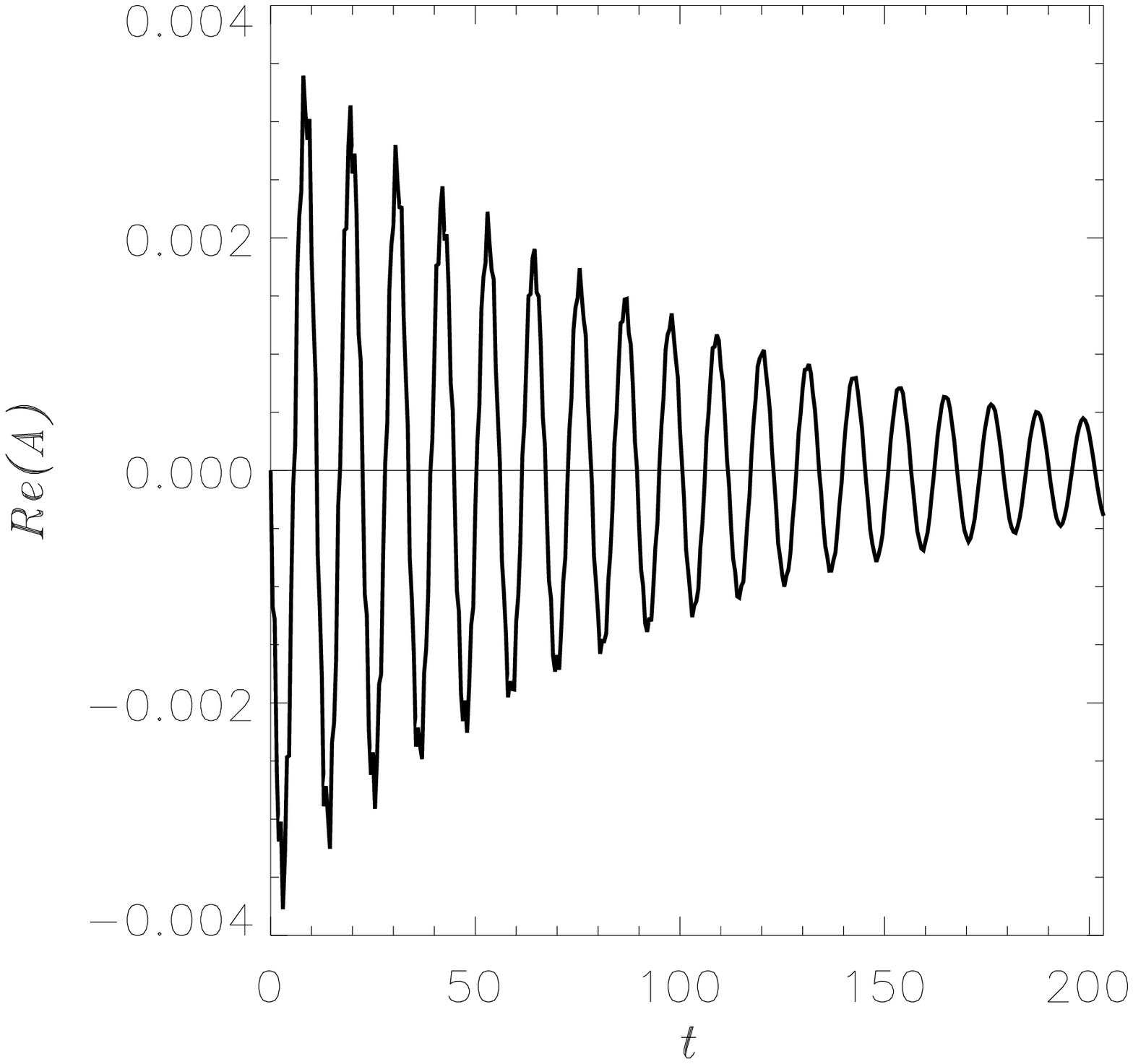}}
%begin mod
 \caption{Left: slide of the velocity field superimposed on the contours
 of entropy perturbations for a 2D-simulation of an oscillating entropy
 bubble embedded in an isothermal atmosphere. Right: time-evolution of
 the real part of its projection coefficient $A_{10}$ (here $t$ is in
 units of $d/c_s$).}
%end mod
\end{figure}

\subsection{Oscillations of an isothermal atmosphere}

As a first test, we apply our method to detect IGW 
%begin mod 
 excited by an oscillating entropy bubble embedded in an isothermal
 atmosphere of depth $d$ (see Fig. 1, left panel). In this case, the building
 of the anelastic subspace is simplified since we found analytic solutions
 for the eigenfrequencies and their associated eigenvectors
 (Dintrans et al., \citeyear{Din02}).
%end mod 

In Fig. 1 (right panel), we show the 
%begin mod 
 real part of the projection coefficient $A_{10}(t)$, i.e. we projected
 the left panel velocity field onto the first anelastic eigenmode
 of the isothermal atmosphere at $k=1$ and $n=0$. As expected, we found
 that $A_{10} (t)$ behaves like $\exp (i\omega_{10}t)$ (with $\omega_{10}
 \simeq 0.569 c_s/d$, where $c_s$ is the constant adiabatic sound speed)
 whereas the mode amplitude $|A_{10}|$ decreases as $\exp (-t/t_\nu)$
 with $t_\nu \propto \nu^{-1}$ ($\nu$ being the constant kinematic
 viscosity of the simulation).
%end mod

\subsection{2D-simulations of penetrative convection}

Once our method validated, we study the excitation of IGW by overshooting
convection using high-resolution two-dimensional simulations of a
three-layer polytropic model. That is, we solve the following equations:

\begin{equation}
\left\{ \begin{array}{l}
\displaystyle \frac{D \Ln \rho}{Dt} = -\Div \vu, \\ \\
\displaystyle \frac{D \vec{u}}{Dt} = - (\gamma-1) \lp \na e + e \na
\Ln \rho \rp + \vec{g} + \frac{1}{\rho} \na \cdot (2\rho \nu \vec{\cal S}),
\\ \\
\displaystyle \frac{De}{Dt} = - (\gamma-1) e \Div \vu +
\frac{1}{\rho} \vec{\nabla} \cdot ({\cal K} \vec{\nabla} e) + \nu
\vec{\cal S}^2 - \frac{e-e_0}{\tau(z)},
\end{array} \right.
\end{equation}
where $\vec{u}$ denotes the velocity, $e$ the internal energy, $\rho$
the density, ${\cal K}=K/c_v$ where $K$ is the radiative conductivity
%begin mod 
 and $c_v$ the specific heat at constant volume, $\vec{\cal S}$ the stress
 tensor and, finally, $\tau(z)$ is a cooling time (see Brandenburg et al.,
 \citeyear{Bra96} for more details).

 Figure 2 (left panel) shows an example of such a simulation of a
 convective zone of depth $d=1$ ($0<z<1$) embedded between two stable
 ones ($-0.15<z<0$ and $1<z<3$). IGW are excited in the bottom radiative
 zone by penetrating downward plumes and the evolution of the projection
 coefficient $A_{10}(t)$ is now more chaotic (see right panel). However,
 by applying a time-frequency diagram on this sequence, we extracted three
 IGW events (with $\omega_{10} \simeq 0.251 \sqrt{g/d}$), emphasized as
 thick lines in the figure.
%end mod 

\begin{figure}
\centerline{\includegraphics[width=0.5\textwidth]{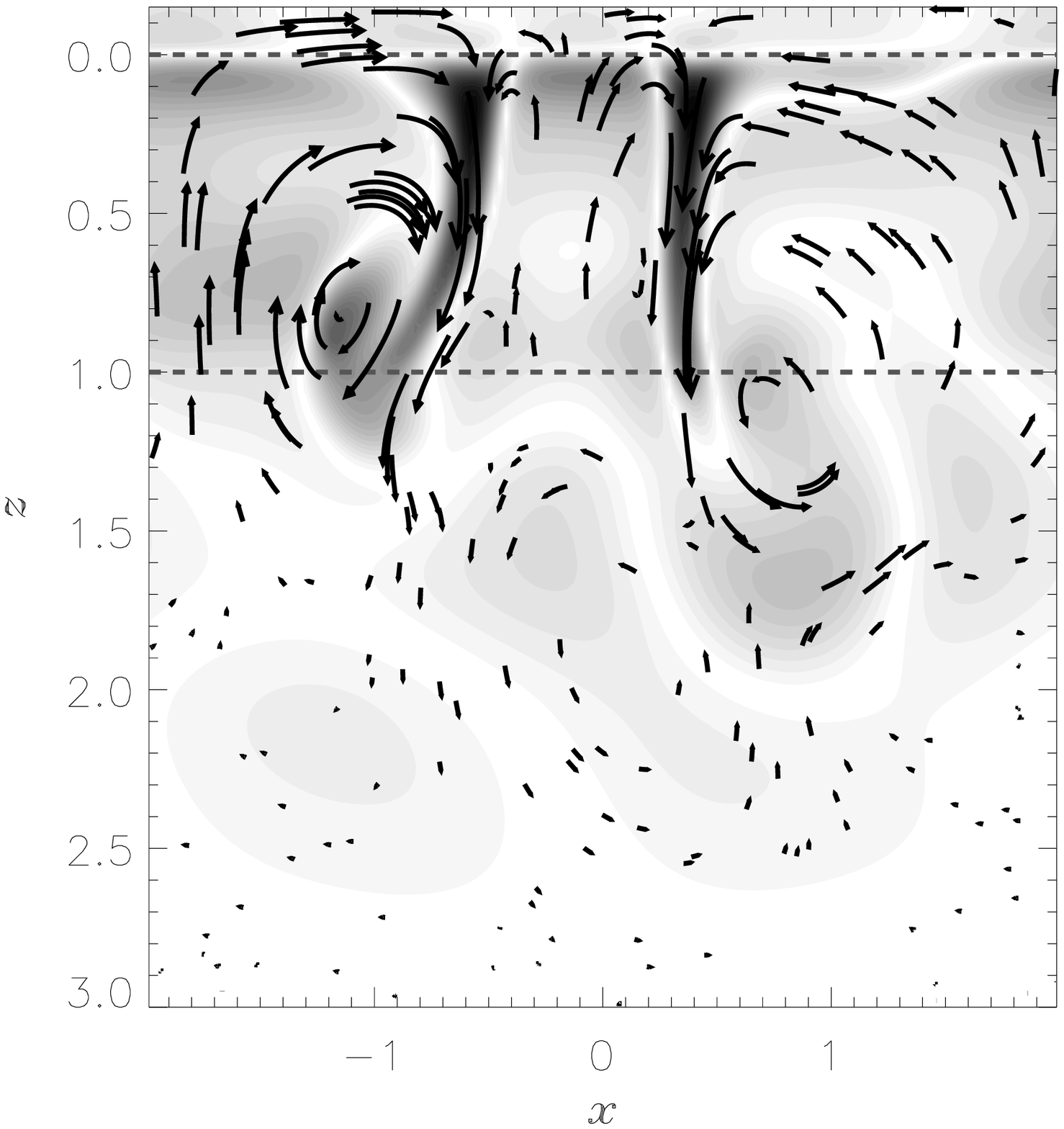}
\includegraphics[width=0.5\textwidth]{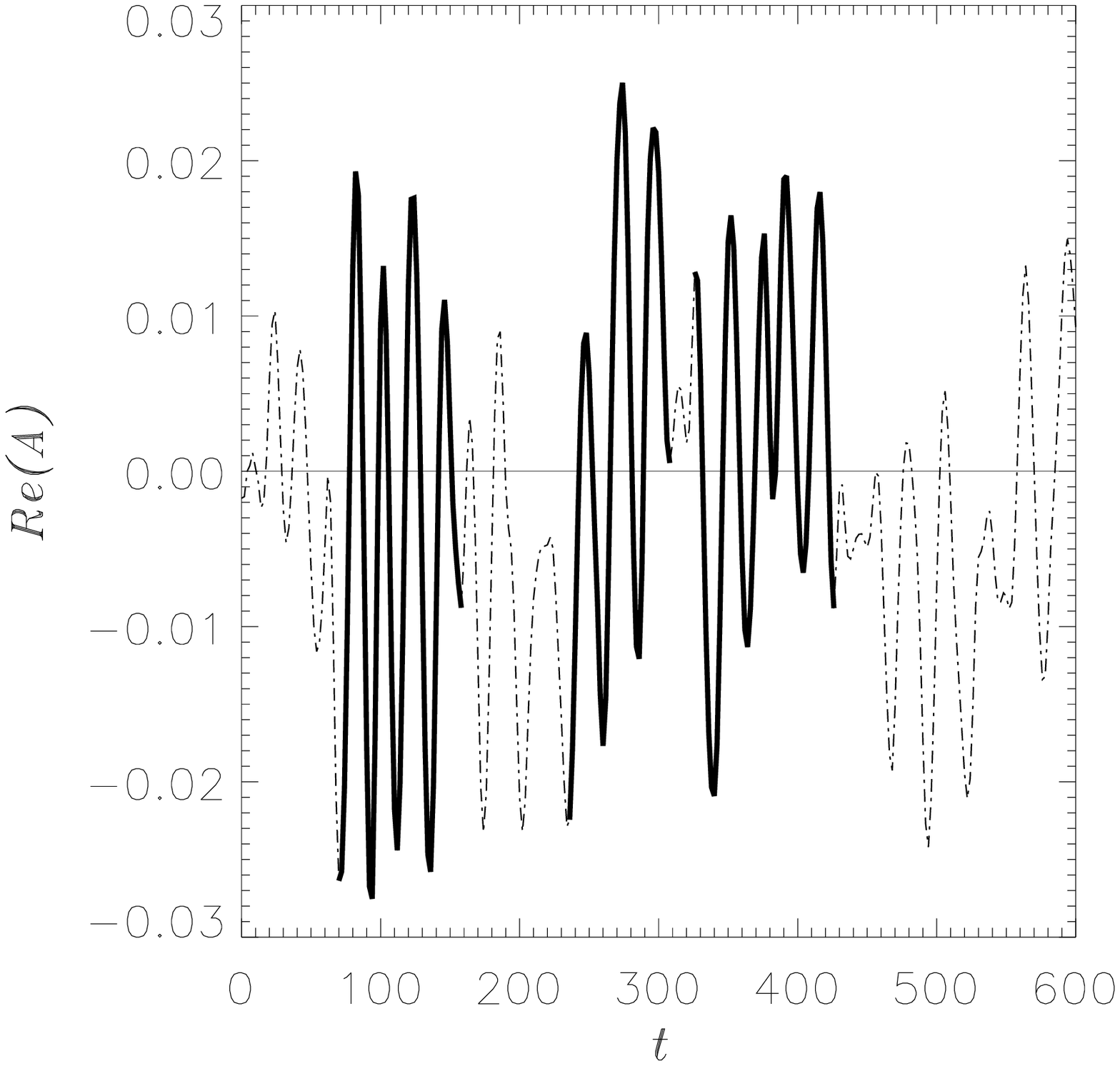}}
%begin mod 
 \caption{Left: slide of the velocity field superimposed on the contours
 of entropy perturbations for a two-dimensional convection simulation. Two
 strong downwards plumes penetrate into the bottom radiative zone and
 excite IGW. Right: time-evolution of the real part of the projection
 coefficient $A_{10}$ (here $t$ is in units of $\sqrt{d/g}$). Thick
 lines emphasize IGW events detected using a time-frequency diagram.}
%end mod
\end{figure}

\acknowledgements
This work has been supported by the European Commission under Marie-Curie
grant no.\ HPMF-CT-1999-00411. Calculations have been carried out on the
CalMip machine of the CICT which is gratefully acknowledged.

%\end{article}
\end{document}